\documentclass[letterpaper]{ESASPCS13Style}
\usepackage{epsfig}
\newcommand{\aapr} {Astron.\&\ Astrophys.\ Rev.}
 \newcommand{\aap}  {Astron.\ \& Astrophys.}
 \newcommand{\aaps} {Astron.\ \& Astrophys.\ Supp.}
 \newcommand{\apjl} {Astrophys.\ J.\ Let.}
 \newcommand{\apj} {Astrophys.\ J.}
 \newcommand{\gca} {Geochim. Cosmochim. Acta}
 \newcommand{\mnras} {Mon.\ Not.\ Roy.\ Astron.\ Soc.}

\newcommand{\aped} {APED}
\newcommand{\asca} {{\it ASCA}}
\newcommand{\chan} {{\it Chandra}}

\newcommand{\heg}  {HEG}
\newcommand{\hetgs} {HETGS}

\newcommand{\letgs} {LETGS}
\newcommand{\meg}  {MEG}
\newcommand{\xmm}  {{\it XMM-Newton}}

\newcommand{\mone}{^{-1}}
\newcommand{\mtwo}{^{-2}}
\newcommand{\mthree}{^{-3}}

\begin{document}
\title{
HETGS Spectroscopy of a Coronally Active Contact Binary, VW Cep
}

\author{
  David P.\ Huenemoerder
 \and  
Adrienne Hunacek
}
\institute{Massachusetts Institute of Technology -- Center for Space
 Research, 77 Massachusetts Avenue, Cambridge, MA 02139, USA
}

\maketitle 

\begin{abstract}
  
  Short-period binaries represent extreme cases in the generation of
  stellar coronae via a rotational dynamo.  Such stars are important
  for probing the origin and nature of coronae in the regimes of rapid
  rotation and activity saturation. VW Cep ($P=0.28$~d) is relatively
  bright, partially eclipsing, and very active object.  Light curves
  made from \chan/\hetgs\ data show flaring and rotational modulation,
  but no strong eclipses. Emission lines are broader than
  instrumental, indicating emission from both binary components.
  Velocity modulation of emission lines is indicative of geometric
  structure of the emitting plasma.

\end{abstract}

\section{Introduction}

\object{VW~Cep} (\object{HD~197433}) is a W-type W~UMa binary --- a
contact binary in which the more massive and larger star has {\em
  lower} mean surface brightness. The primary (deeper) photometric
eclipse is the occultation of the smaller star.  VW~Cep has an 0.28 d
(24 ks) orbital period.  It is partially eclipsing ($i=63^\circ$),
with component spectral types of K0~V and G5~V
(\cite{Hill:1989}, \cite{Hendry:Mochnacki:2000}).  It is one of the
X-ray-brightest of contact binaries.

Among the coronally active binaries, activity is strongly correlated
with rotation (\cite{Pallavicini:1989}), but saturates at periods
below one day (\cite{Vilhu:Rucinski:1983}, \cite{Cruddace:Dupree:1984}).  The
trend has been referred to as ``super-saturation'' since instead of
reaching a plateau, the activity level actually decreases with
increasing rotation rate
(\cite{Randich:1998}, \cite{Jardine:Unruh:1999},
\cite{James:Jardine:al:2000}, \cite{Stepien:Schmitt:Voges:2001}).
\cite{Jardine:Unruh:1999} argued that super-saturation occurs because
loops are large and are unstable to the coriolis forces as they exceed
the co-rotation radius: extended loops are swept to the poles.  On the
other hand, \cite{Stepien:Schmitt:Voges:2001} contended that loops are
quite compact (relative to the stellar radius), but that the dynamics
of surface flows in contact systems conspire to clear equatorial
regions.  The two scenarios are similar, but they differ in an
important respect: X-ray sources are either low volume and dense, or
large volume and rarefied.  Optical light curve modeling is consistent
with either (assuming correlation between photospheric spots and
coronal emission); Doppler image maps of VW~Cep
(\cite{Hendry:Mochnacki:2000}) showed large polar spots.

\cite{Choi:Dotani:1998} have analyzed \asca\ spectra, and found a flux
of about $1\times10^{-11}\,\mathrm{ergs\,cm^{-2}\,s^{-1}}$, and two
component model temperatures of $7$ and $22\times10^6$K ($\log T\sim
6.8$ and $7.3$) with about equal emission measures.  They obtained
significantly reduced abundances of Fe, Si, Mg, and O, but a Solar
value for Ne.  A flare also occurred during the observation, with a
factor of three increase in the count rate.  They used the flare
emission measure and loop-scaling models to estimate a density of
about $5\times10^{10}\mathrm{cm^{-3}}$.  The Ginga observations
(\cite{Tsuru:al:1992}) showed a thermal plasma temperature in excess
of $10^8\,$K, a flux similar to that determined by
\cite{Choi:Dotani:1998}, no rotational modulation, and Fe~K flux lower
than expected.

\section{High Resolution X-Ray Spectra}

We observed VW~Cep for 116~ks with the \chan\ High Energy Transmission
Grating Spectrometer (\hetgs) on August 29-30, 2003 (observation
identifier 3766).  The instrument has a resolving power ($E/\Delta E$)
of up to 1000 and wavelength coverage from about 1.5~\AA\ to 26~\AA\ 
in two independent channels, the High Energy Grating (\heg), and
Medium Energy Grating (\meg).

VW~Cep has also been observed in X-rays with the Chandra Low Energy
Transmission Grating Spectrometer (\cite{Hoogerwerf:al:2003}), and
with the \xmm\ X-ray observatory (\cite{Gondoin:2004}).  The only
other contact binary observed with \hetgs\ is 44~Boo
(\cite{Brickhouse:01}).  Observations with the \hetgs\ provide
spectral resolution and sensitivity required to better determine line
fluxes, wavelengths, and profiles.  Ultimately, we wish to determine
which coronal characteristics are truly dependent upon fundamental
stellar parameters.  Combination of VW~Cep data with that from \letgs\ 
and \xmm, as well as with results for other short period systems
(e.g., 44~Boo, ER~Vul) will help us to understand the basic emission
mechanisms.

Here we concentrate on preliminary HETGS results.  Data were
calibrated with standard CIAO pipeline and response tools\footnote{\tt
  http://cxc.harvard.edu/ciao}. Further measurements and analysis were done
in ISIS (\cite{Houck:00}) and with custom code using ISIS as a
development platform.
\begin{figure*}[!htb]
  \begin{center}
    \epsfig{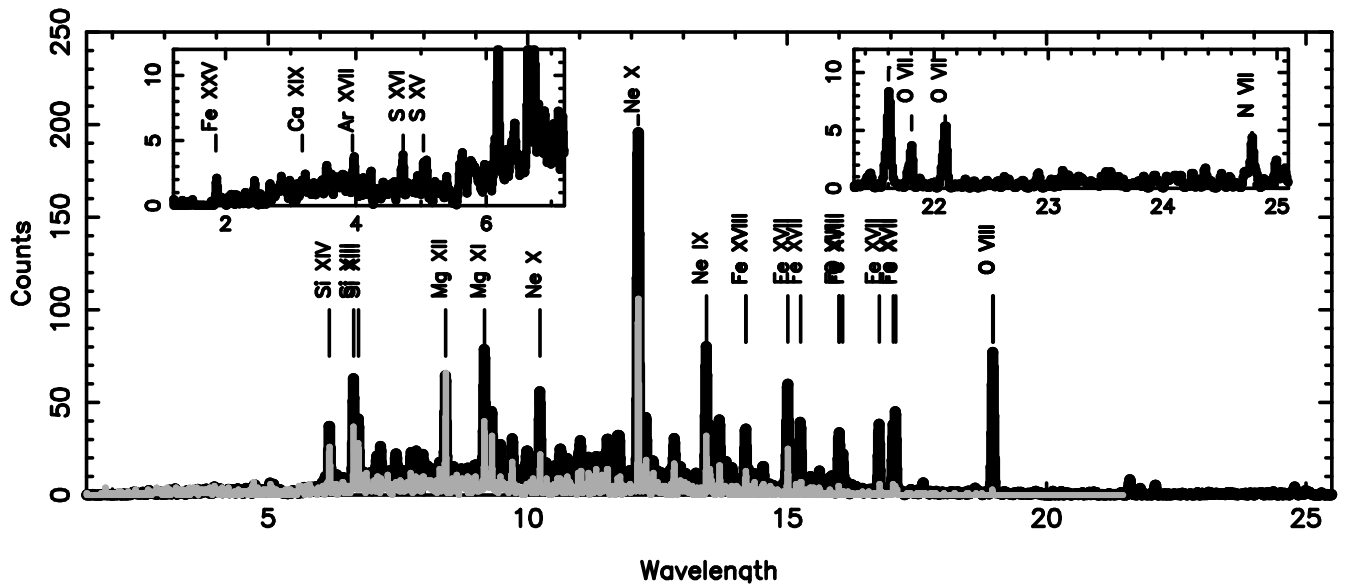}
  \end{center}
  \caption{HETGS spectrum of VW~Cep, 116 ks exposure.  MEG (16433
    counts) is the black line, and HEG (5269 counts) the gray.  Insets
    show detail of the HEG spectrum (left) and MEG O~VII triplet and
    N~VII region (right).
    \label{fig:spec}
  }
\end{figure*}

\section{The Spectrum}

Figure~\ref{fig:spec} shows the counts spectrum obtained from the
HETGS full exposure.  The spectrum is qualitatively typical of coronal
sources: a variety of emission lines from highly-ionized elements
formed over a broad temperature region, from \ion{O}{vii},
\ion{N}{vii}, \ion{Ne}{ix}, and \ion{Fe}{xvii} ($\log T\sim
6.3$--$6.7$), up to high temperature species like \ion{S}{xv},
\ion{S}{xvi}, \ion{Ca}{xix} (perhaps), and \ion{Fe}{xxv} ($\log T\sim
7.2$--$7.8$).  It is also apparent that iron has a fairly low
abundance relative to neon, given the relative weakness of the 15\AA\ 
and 17\AA\ \ion{Fe}{xvii} lines relative to the \ion{Ne}{ix} 13\AA\ 
lines.  The observed flux in the 2--20~\AA\ is
$7\times10^{-12}\,\mathrm{ergs\,cm\mtwo\,s\mone}$, and the luminosity
(for a distance of 27.65 pc) is
$6.5\times10^{29}\,\mathrm{ergs\,s\mone}$.

Cursory inspection of the density-sensitive helium-like triplet lines
(\ion{O}{vii}, \ion{Ne}{ix}, \ion{Mg}{xi}) does not reveal ratios
obviously far from their respective low-density limits.

\section{Emission Measure}

The differential emission measure ($DEM$) is a one dimensional
characterization of a plasma, and can be defined as $N_e^2\,dV/dT$, in
which $N_e$ is the electron density, $V$ is the emitting volume, and
$T$ the temperature.  The $DEM$ is an important quantity because it
represents the radiative loss portion of the underlying heating
mechanism.  An emission measure can be derived from measurements of
line fluxes and some assumptions about the homogeneity and ionization
balance of the emitting plasma.  Derivation of the emission measure
thus relies on detailed knowledge of fundamental atomic parameters.
Even given perfect knowledge of the ionic emissivities, their
contribution functions versus temperature are broad so the emission
integral cannot be formally inverted.  Hence there are many methods to
regularize the problem and solve for the emission measure and
elemental abundances.  For a first estimate, we use the simple method
described by \cite{Pottasch:1963}, in which the $DEM$ is approximated
by a ratio of line luminosity to average line emissivity at the
temperature of the maximum emissivity.  Since there is a degeneracy in
elemental abundance and $DEM$, the relative abundance of one element
is scaled to bring their $DEM$ into better agreement with the locus of
some other element with peak emissivities near the same temperatures
(typically Fe, since it has many ions over a broad range in
temperature). 

\begin{figure}[!ht]
  \begin{center}
    \epsfig{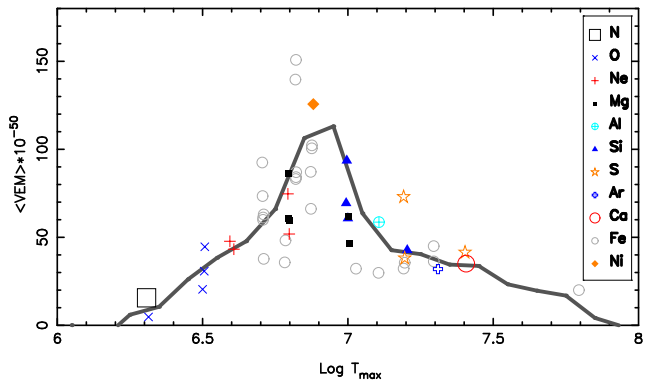}
  \end{center}
  \caption{A provisional emission measure constructed from the ratio
    of line fluxes to mean emissivity over 0.3 dex.  Points are
    plotted for each line used, and different elements points were
    scaled as a group to be roughly consistent with others.  The solid
    line is a smoothed average of the individual points.  The
    integrated emission measure is about
    $6.8\times10^{52}\,\mathrm{cm\mthree}$. 
    \label{fig:dem}
  }
\end{figure}
Figure~\ref{fig:dem} shows our {\em provisional} $DEM$,
using the Astrophysical Plasma Emission Database (\aped) for line
emissivities (\cite{Smith:01}), the ionization balance of
\cite{Mazzotta:98}, and Solar abundances of \cite{Anders:89}.  There
is large scatter in the values; the solid curve shows a smoothed
average.  There is sharp structure, with a fairly large peak at $\log
T=6.9$.  This $DEM$ can be used to generate a synthetic spectrum for
further refinement of a plasma model.  We found that with this
distribution, we required a neon abundance of 1.0 (relative to Solar),
and 0.3 for iron, confirming a trend already inferred from visual
inspection of the spectrum.  In iterative $DEM$ reconstruction
techniques, sharper structure can be modeled than in the simple
method.  The fact that the simple method shows a strong peak hints at
a highly structured $DEM$ in VW~Cep.

\section{Light and Phase Curves}

\begin{figure}[!ht]
  \begin{center}
    \epsfig{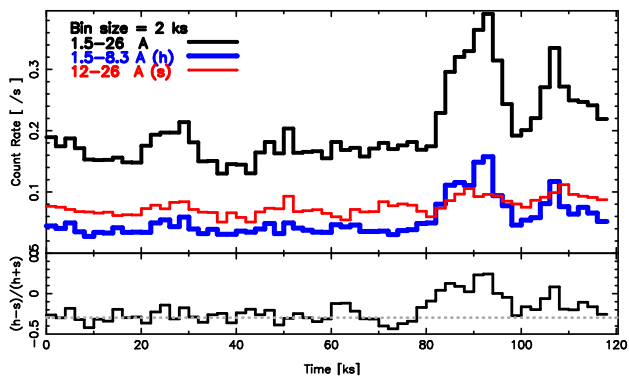}
  \end{center}
  \caption{Count rates of VW~Cep in 2 ks bins.  In the upper panel
    are light curves extracted from all diffracted photons (1.5-26\AA;
    upper heavy (black) curve), a ``soft'' band (12-26\AA; lower, thinner
    curve (red) ),
    and a ``hard'' band (1.5-8.3\AA; lower, thicker curve (blue)).
    The lower panel shows a hardness ratio (solid histogram), and the
    median of the hardness for the first 70 ks (light dotted line).
    \label{fig:lc}
  }
\end{figure}

Given the short orbital period, the observation covered almost five
revolutions without interruption.  Such phase redundancy is important
to discriminate intrinsic variability from that caused by
rotational or eclipse modulation.  Figure~\ref{fig:lc} shows light
curves in several wavelength regions and a hardness ratio.  There was
much variability.  The hardness shows that the large increase near
after 80 ks is a flare, by definition of being hotter: proportionally
more flux was emitted at shorter wavelengths, which are very sensitive
to high temperatures via the thermal continuum emission.  Conversely,
the bump in count rate between 20-30 ks does not show in hardness, and
must be due to rotational modulation.

\begin{figure}[!t]
  \begin{center}
    \epsfig{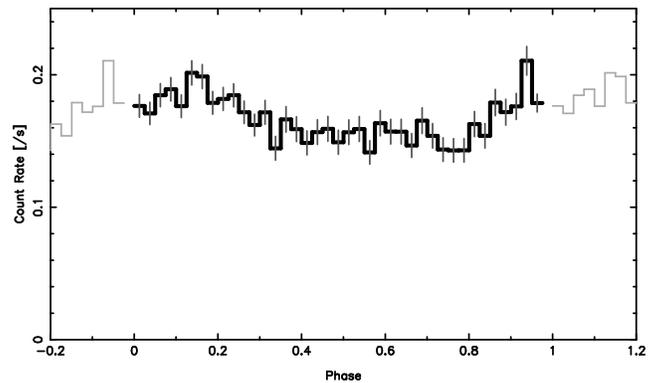}
  \end{center}
  \caption{The phase-folded X-ray count-rate curve of VW~Cep,
    excluding flare times.  Bins are $\Delta\phi=0.025$, or 600 s,
    with a cumulative exposure per bin of 1800 to 2400 s, depending on
    the phase.
    \label{fig:pc}
  }
\end{figure}

If we exclude the flare times and phase-fold the remaining data (about
3 orbital periods), we easily see the rotational modulation.
Figure~\ref{fig:pc} shows this curve, using the ephemeris of
\cite{Pribulla:al:2002}.  
No eclipses are apparent, but instead, quasi-sinusoidal modulation of
about 18\%.  The visual light curve (\cite{Pribulla:al:2000}) shows a
similar amplitude (16\%), but is very different qualitatively, with
strong minima at phases 0.0 and 0.5, and continuously variable in
between (a trademark of W~UMa systems).  Without additional
information, the X-ray light curve is not sufficient for localizing
emission to one star or another.  All we can say is that there is some
asymmetric distribution, and possible occultation of large structures.

\section{ Velocity Modulation }

\begin{figure}[!ht]
  \begin{center}
    \epsfig{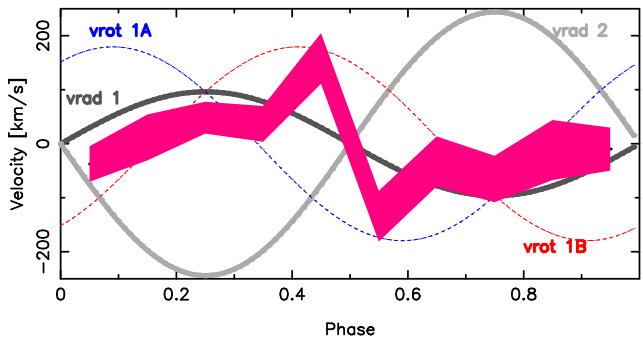}
  \end{center}
  \caption{Mean wavelengths of photons in the \ion{Ne}{x} (12.1\AA)
    region of the spectrum.  The shaded area is the one-sigma
    envelope.  Sinusoids are either radial velocities of the stellar
    centers of mass (black heavy curve: primary, ``vrad1''; gray heavy curve:
    secondary, ``vrad2''), or of points on opposing hemispheres of the
    primary (``vrot1A'' and ``vrot1B'').  Sinusoids are approximate;
    they do not account for the tidal distortions of contact binaries.
    \label{fig:nex}
  }
\end{figure}
In the high-resolution spectrum, additional information is available
in the line shapes and centroids.  The lines of VW~Cep are broader
than the instrumental resolution when integrated over the entire
observation.  This means that we are most likely sensitive to orbital
Doppler shifts of one or both components, or to rotational broadening.

In order to investigate further, we have measured the mean wavelength
of events in the best exposed line, \ion{Ne}{x} (12.1\AA), in phase
bins of 0.1 (also excluding flare times).  The one-sigma envelope is
shown in Figure~\ref{fig:nex}.  It roughly follows the radial velocity
of the primary (more massive) star (heavy dark sinusoid), except near
phase 0.5, where it does a flip-flop, moving red-ward, then blue-ward.
Our current hypothesis is that we are seeing a bias in the centroid
during the transit of the secondary which first blocks the approaching
limb of the primary, then the receding limb as the transit progresses.
The dashed sinusoidal curves in the figure denote approximate
photospheric rotational velocities on opposite sides of the primary.
The velocity amplitude of a point on the stellar surface is consistent
with the observed perturbation.

\section{Conclusions and Future Work}

The velocity curve of \ion{Ne}{x} suggests that the emission is
predominantly associated with the primary (larger, more massive)
stellar component.  This is also consistent with coincidence of the
amplitude of the count-rate curve with the optical amplitude, and with
the lack of a distinct primary eclipse (smaller, secondary star
occulted).  The velocity flip-flop perturbation also suggests that
material is near the equatorial radius --- either compact and near the
equator itself, or extended at mid-latitudes, but co-rotating.

If the dip in rate seen by \cite{Gondoin:2004} in \xmm\ data taken
only 10 months earlier is really an eclipse, then this indicates
rather large and rapid changes in coronal structure.  The \xmm\
observation, however, did not cover a full continuous period, and the
dip is offset from phase 0.0, so the interpretation is difficult.

The line spectrum and crude $DEM$ reconstruction show that the the
emission comes from a range of temperatures, and that the emission
measure is highly structured.  The relative abundances of Ne and Fe
also appear to be similar to other coronal sources: Ne enhanced by a
factor of a few relative to Fe.

In the future, we will perform a $DEM$ and abundance analysis, in
which the line-based simultaneous $DEM$ and abundance solutions will
be used to predict a better model continuum, and thereby improve line
flux measurements, which are then used to iterate the model.  We will
also perform analysis of composite line profiles in order to refine
the velocity versus phase analysis.  By using a dozen or so lines,
velocity sensitivity can be reduced to about $30\,\mathrm{km\,s\mone}$
(see \cite{Hoogerwerf:al:2004}, for example).  We will perform some
analyses of the flare data separately to search for line shifts or
flare occultation.  And finally, we will do careful measurement and
modeling of the density-sensitive He-like triplet ratios.

\begin{acknowledgements}
  This research was supported by NASA grant G03-4005A and SAO contract
  SV1-61010.
\end{acknowledgements}

%

\end{document}